\documentclass[journal]{IEEEtran}
\usepackage{cite}

\ifCLASSINFOpdf
   \usepackage[pdftex]{graphicx}
\else
\fi
\usepackage{amsmath}
\usepackage{amssymb}
\usepackage{bm}
\usepackage{cases}
\usepackage{array}
\usepackage{url}

\begin{document}
%
\title{Round-core-radius Dependent Electromagnetic Coupling of Multi-filament Helical Superconducting Tapes in a Swept Magnetic Field}
%
%
%

\author{
Yoichi~Higashi~and~Yasunori~Mawatari
\thanks{Y. Higashi and Y. Mawatari are with National Institute of Advanced Industrial Science and Technology (AIST), Tsukuba, Ibaraki 305-8568, Japan (e-mail: y.higashi@aist.go.jp).}
\thanks{Manuscript received mmmm dd, yyyy; revised mmmm dd, yyyy.}
}

%
%

\markboth{Journal of \LaTeX\ Class Files,~Vol.~14, No.~8, August~2015}%
{Shell \MakeLowercase{\textit{et al.}}: Bare Demo of IEEEtran.cls for IEEE Journals}
%



\maketitle

\begin{abstract}
With the excitation and demagnetization of a magnet for magnetic resonance imaging in mind,
we theoretically and numerically investigated electromagnetic coupling---especially its dependence on the round-core radius $R$---of multi-filament helically wound superconducting tapes
under steady-state conditions 
in a constantly ramped magnetic field.
We found that even in a rapidly ramped magnetic field,
the electromagnetic coupling can be suppressed by reducing $R$ to close to the tape width.
We also clarified that the coupling sweep rate at which the electromagnetic coupling starts
scales as $R^{-2}$,
showing that the dependence on $R$ reflects the penetration of magnetic flux from the edges of the tape.
Even when the round core is as narrow as the tape width,
the behavior is considered to be similar to the electromagnetic response of a flat tape rather than that of a tubular wire.
\end{abstract}


%
\IEEEpeerreviewmaketitle

\section{Introduction}
\IEEEPARstart{C}{}ORC (Conductor on Round Core) wire, 
which comprises a rare-earth barium copper oxide YBa$_2$Cu$_3$O$_{7-\delta}$ coated conductor on a round core \cite{laan2009,weiss2017},
is a promising candidate for superconducting (SC) wires for applications involving high-field magnets.
Among all high-temperature SC cables and wires, CORC ones have the shortest pitch lengths, as short as several tens of millimeters.
As such, the electromagnetic (EM) coupling between SC filaments through a plated thermally stabilizing layer is expected to be suppressed,
and striation can facilitate further loss reduction.
Some previous studies fabricated CORC cables made of striated coated conductors and evaluated their alternating-current losses experimentally \cite{souc2013,vojenciak2015},
but the SC filaments were completely separated by striation; that is, they were insulated. 
Regarding local heating in the SC filaments,
a copper stabilizing layer is coated on top of the multi-filament tapes for increased thermal stability. 
Very recently, a spirally twisted multi-filament coated conductor with finite transverse conductance between the SC filaments was fabricated,
and measurements were made regarding how the magnetization losses depended on the frequency of a sinusoidally oscillating applied field \cite{li2020}.
At low transverse field amplitudes,
selective measurements were made of the dominant contribution from coupling losses to the magnetization loss
to determine the coupling time constants \cite{li2020}.

Regarding theoretical studies,
based on the thin-sheet approximation using the current vector potential,
EM analyses have been performed to investigate the magnetization loss and EM coupling in a sinusoidally oscillating applied field \cite{wang2019,sogabe2020,yan2020}.
Numerical simulation has shown considerable reduction of the magnetization loss by striation in high magnetic fields,
although an increase was seen in the magnetization loss in low fields when there were only a few filaments \cite{wang2019}.

Despite the aforementioned studies, how the EM coupling depends on the core radius remains unclear and is yet to be investigated systematically.
Clarifying the conditions for the EM coupling would give valuable information about how fast an applied field can be swept
given the practical parameters of the CORC cable or wire.
As opposed to previous theoretical studies \cite{wang2019,sogabe2020,yan2020},
we consider a swept magnetic field with the excitation and demagnetization of a magnetic resonance imaging (MRI) magnet in mind.

In the present study, we use numerical simulations to investigate the magnetization loss and EM coupling
of a multi-filament helical SC tape in an applied magnetic field with a high sweep rate of $\beta=300$~mT/s,
which is two orders of magnitude higher than a typical sweep rate of an MRI magnet; 
in particular, we focus on the dependence on the core radius.
We assume, for example, the fast demagnetization of an MRI magnet during an emergency stop.

\section{Model of Multi-filament Helical Superconducting Tape}
\begin{figure}[tb]
\centering
\includegraphics[width=3.5in]{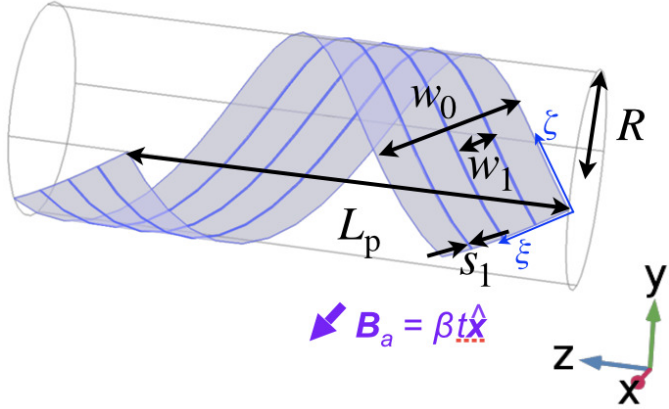}
\caption{Schematic of four-filament helically wound superconducting (SC) tape of total width $w_0$, filament width $w_1$, resistive slot width $s_1$, and pitch $L_{\rm p}$ on a hollow cylinder of radius $R$.
The axes $(\xi,\zeta)$ are on the tape surface.
}
\label{fig0}
\end{figure}
We examine numerically the model of a four-filament helical SC tape with conductors of finite normal resistivity $\rho_{\rm n}$ embedded in slots between the filaments.
We assume that the coupling current between the SC filaments flows uniformly through the thermal stabilizing layer.
The stabilizer information (e.g., conductivity, thickness, and effective length responsible for the transverse conductance)
can be incorporated via the transverse conductivity (resistivity) $\sigma_{\rm n}=1/\rho_{\rm n}$ as discussed in Section~\ref{sec3},
thereby reducing a three-dimensional model to a two-dimensional model \cite{amemiya2018}.
Here, we model a thermal stabilizer such as copper on top of the SC filaments by means of a finite-resistivity conductor as in Refs.\ \cite{amemiya2006,amemiya2018,higashi2019_sust}.
Fig.~\ref{fig0} shows the four-filament helical SC tape schematically.
Unless specified otherwise, the dimensions are as follows: the total tape width is $w_0=2$~mm, the width of an SC filament is $w_1=0.485$~mm, the slot width is $s_1=20~\mu$m,
the thickness of the SC tape is $d_0=2~\mu$m, and the radius of the hollow cylinder is $R=3$~mm.

The helical SC tape is assumed to be so thin ($d_0 \ll w_0$) that it 
can be approximated as an infinitesimally thick surface described by the coordinates \cite{higashi2020}
\begin{equation}
\label{helicoid-coordinate}
\setlength{\nulldelimiterspace}{0pt}
\left\{
\begin{IEEEeqnarraybox}[\relax][c]{l's}
x=R \cos[ (k\zeta-\xi/R )/\sqrt{1+(kR)^2} ],\\
y=R \sin[(k\zeta-\xi/R)/\sqrt{1+(kR)^2}],\\
z=(\zeta+kR\xi)/\sqrt{1+(kR)^2},
\end{IEEEeqnarraybox}
\right.
\end{equation}
where $k=2\pi/L_{\rm p}$ and $L_{\rm p}$ is the pitch length of the helical conductor.
The axial coordinates $(\xi,\zeta)$ are orthogonal on the tape surface.
We consider the helical SC tape surface with a full conductor pitch $L_{\rm p}$ spanned by $-w_0/2\le \xi \le w_0/2$ and $-L_{\rm tape}/2\le \zeta \le L_{\rm tape}/2$,
with $L_{\rm tape}= L_{\rm p}\sqrt{1+(k R)^2}$ being the tape length (see Figs.~\ref{fig0} and \ref{fig1}).

\section{Response of Multi-filament Helical Tape to a Ramped Field}
\label{sec3}
Because an MRI magnet operates in a magnetic field as high as a few teslas,
we may assume that the parts of the SC tape that are perpendicular to the applied field are fully penetrated by magnetic flux vortices.
For a concrete estimate,
in a multi-filament helical tape,
the full penetration field can be estimated using the formula for a flat tape, namely, 
$B_{\rm p}=(\mu_0 J_{\rm c} d_0/\pi)\left[ 1+\ln(w_1/d_0)\right]\approx 0.260$~T \cite{brandt1996},
where $\mu_0$ is the magnetic permeability of a vacuum and $J_{\rm c}$ is the critical current density. Meanwhile,
in the parts of the tape that are parallel to the applied field,
the perpendicular field component is small and does not reach $B_{\rm p}$,
and the loss generated in such parts is negligibly small.

In the fully penetrated state,
the spatial profile of the magnetic field on the SC tape does not change,
being kept in the steady state,
as long as we disregard the magnetic field and its angle dependence of $J_{\rm c}$;
that is, $J_{\rm c}$ is constant.
The approximation of constant $J_{\rm c}$ is effective for evaluating the power loss \cite{higashi2019}.

At high fields, magnetization losses dominate transport losses \cite{kajikawa2016},
and so we neglect the contribution from transport losses in the present study.
We also neglect the magnetic field due to the screening current 
because it is much smaller than the applied field.

In previous work, by adopting the thin-sheet approximation that takes account of only the perpendicular field component,
we derived the steady-state Faraday--Maxwell equation for a constantly swept applied magnetic field
$\bm{B}_{\rm a}=\beta t \hat{\bm{x}}$
with rate $\beta={\rm d}B_{\rm a}/{\rm d}t$ \cite{higashi2019,higashi2020}.
We base the present numerical simulation on the following two-dimensional Faraday--Maxwell equation on the tape surface $(\xi,\zeta)$
extended to multi-filament helical SC tapes:
\begin{equation}
\frac{\partial}{\partial \xi}\left( \rho \frac{\partial g}{\partial \xi} \right) + \frac{\partial}{\partial \zeta}\left( \rho \frac{\partial g}{\partial \zeta}\right)
=\beta \cos\left[\frac{k\zeta-\xi/R}{\sqrt{1+k^2R^2}}\right],
\label{faraday-maxwell-eq}
\end{equation}
where $g(\xi,\zeta)$ a scalar function defined on the helical SC tape surface.
The contour lines of $g(\xi,\zeta)$ describe the current streamlines on the helical tape surface (see Fig.~\ref{fig1}).
 
To model a multi-filament helical SC tape, we adopt the SC nonlinear resistivity $\rho_{\rm sc}$ and the transverse linear one $\rho_{\rm n}$ for the filaments and slots, respectively, namely,
\begin{equation}
\rho(\xi,\zeta)=
\setlength{\nulldelimiterspace}{0pt}
\left\{
\begin{IEEEeqnarraybox}[\relax][c]{l's}
\frac{E_{\rm c}}{J_{\rm c}}\left( \frac{|\bm{J}(\xi,\zeta)|}{J_{\rm c}} \right)^{n-1}\equiv \rho_{\rm sc}(|\bm{J}|)~~{\rm for~SC~filaments},\\
\rho_{\rm n}~~{\rm for~resistive~slots},
\end{IEEEeqnarraybox}
\right.
\end{equation}
which is defined via the electric field $\bm{E}$--current density $\bm{J}$ characteristic of $\bm{E}=\rho\bm{J}$.
Herein, we disregard the dependence of $J_{\rm c}$ on the the magnetic field and its angle.
We set 
$J_{\rm c}=5\times10^{10}$~A/m$^2$, $E_{\rm c}=1$~$\mu$V/cm for the electric field, and $n=15$ in the power of the SC nonlinear resistivity. 
We incorporate the effect of a copper thermal stabilizer on the multi-filament helical SC tape through $\rho_{\rm n}$ \cite{higashi2019_sust,amemiya2018}.
The transverse conductance is obtained as $G_{\rm n}=\sigma_{\rm n}(L_{\rm tape}/2)d_0/s_1$, where we use the effective tape length $L_{\rm tape}/2$ (see Fig.~\ref{fig1}).
Consequently, the transverse conductance per unit length is $G_{\rm t}=G_{\rm n}/(L_{\rm tape}/2)=\sigma_{\rm n}d_0/s_1$.
When the coupling current flows mainly through the copper stabilizer plated on the multi-filament helical tape,
we can equate $G_{\rm t}$ with the conductance per unit length via the plated copper stabilizer, namely, $G_{\rm Cu}=\sigma_{\rm Cu}d_{\rm Cu}/l_{\rm Cu}$,
where $\sigma_{\rm Cu}$ is the conductivity of copper, $d_{\rm Cu}$ is the thickness of the plated copper, and $l_{\rm Cu}$ is the effective length responsible for the transverse conductance via the plated copper layer. 
Therefore, by adopting $d_{\rm Cu}= 20~\mu$m, $l_{\rm Cu}=300~\mu$m \cite{amemiya2018}, and $\sigma_{\rm Cu}=5\times10^8~{\rm \Omega}^{-1}$m$^{-1}$ at 77~K \cite{hucek1977handbook},
we have that
\begin{equation}
\sigma_{\rm n}=\frac{s_1}{d_0}\frac{d_{\rm Cu}}{l_{\rm Cu}}\sigma_{\rm Cu}\approx 3.33 \times10^8~{\rm \Omega}^{-1}{\rm m}^{-1},
\end{equation}
which yields $\rho_{\rm n}\sim1\times10^{-9}$ $\Omega$m, and we use this value throughout the paper.

\section{Numerical Results}

\subsection{Current Streamlines}
\begin{figure*}[tb]
\centering
\includegraphics[width=7.3in]{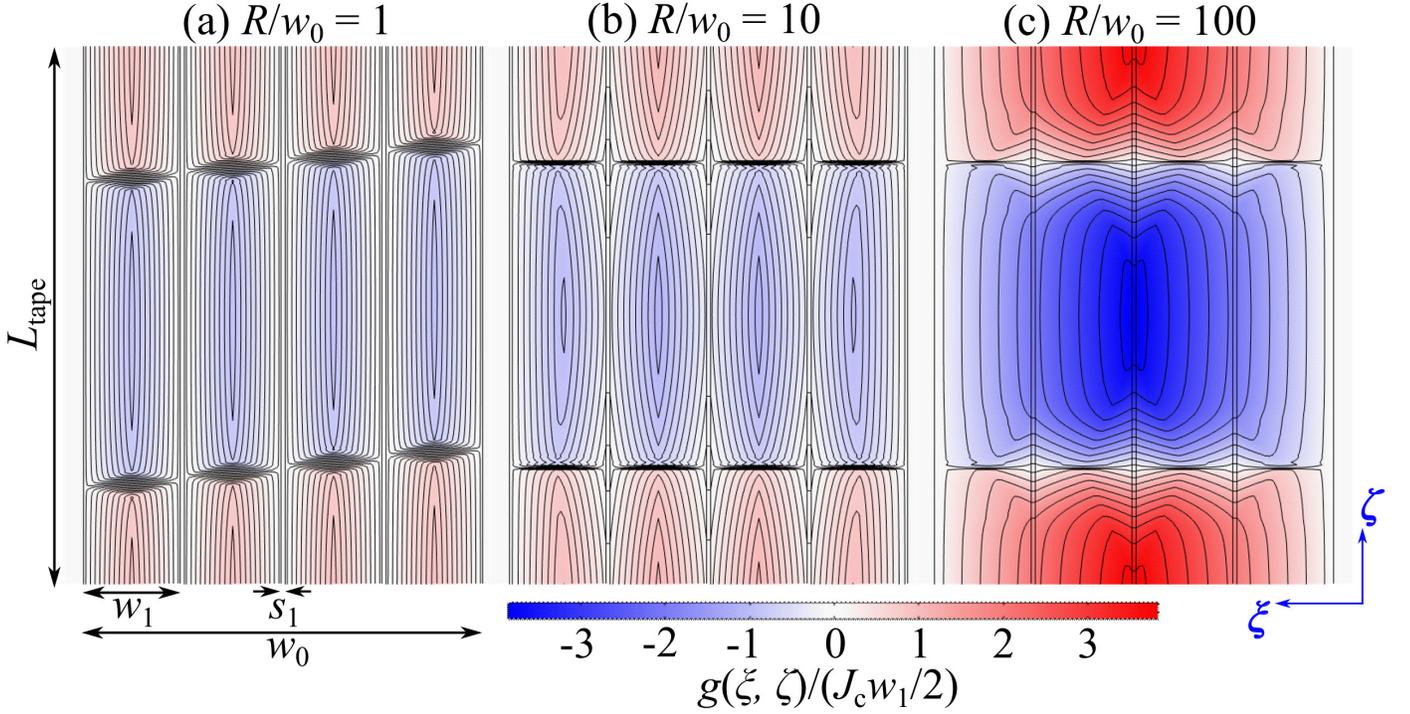}
\caption{Current streamlines and spatial profile of $g(\xi,\zeta)$ on surface of multi-filament helical SC tape
for $R/w_0=$ (a) 1, (b) 10, and (c) 100.
The pitch length and sweep rate are $L_{\rm p}=6$~mm and $\beta=300$~mT/s, respectively, and 
the tape length for a full pitch is $L_{\rm tape}\approx$ (a) 13.9, (b) 125.8, and (c) 1256.7~mm.
}
\label{fig1}
\end{figure*}
To obtain the solution for $g(\xi,\zeta)$, we solve Eq.~(\ref{faraday-maxwell-eq}) numerically using the commercial software COMSOL Multiphysics\textsuperscript{\textregistered} \cite{comsol}.
We impose the Dirichlet boundary condition $g(\xi=\pm w_0/2,\zeta)=0$ along the long edges of the tape, 
and we impose the periodic boundary condition
\begin{eqnarray}
g(\xi, \zeta=-L_{\rm tape}/2)&=&g(\xi,\zeta=L_{\rm tape}/2)
\end{eqnarray}
on the terminals of the whole tape.
The $\xi$ and $\zeta$ components of the current density on the tape surface are obtained from the spatial derivatives of $g(\xi,\zeta)$ as \cite{higashi2020}
\begin{equation}
J_\xi=-\frac{\partial g(\xi,\zeta)}{\partial \zeta},~~J_\zeta=\frac{\partial g(\xi,\zeta)}{\partial \xi}.
\end{equation}
In Fig.~\ref{fig1}, the solid lines depict the current streamlines on the tape surface, and 
the color scale shows the profile of the scalar function $g(\xi,\zeta)/(J_{\rm c}w_1/2)$.
At $R/w_0=1$ [Fig.~\ref{fig1}(a)], each SC filament is completely decoupled electrically.
However, partial EM coupling appears at $R/w_0=10$ [Fig.~\ref{fig1}(b)]
because the current streamlines go partially across the resistive slots.
At $R/w_0=100$  [Fig.~\ref{fig1}(c)],
the current streamlines spread throughout the tape surface,
and thus the SC filaments are completely coupled electrically.
As shown in Fig.~\ref{fig1}, the helical winding divides the current streamlines into closed loops for every half pitch,
showing that the effective tape length is $L_{\rm tape}/2$.
However, unlike for multi-filament twisted tapes \cite{higashi2019_sust},
note that for the parameters adopted in the present study, one cannot obtain the correct solution for $g(\xi,\zeta)$ at least when $R/w_0=1$
if the Dirichlet boundary condition $g(\xi,\zeta=\pm L_{\rm tape}/4)=0$
is imposed where the tape is parallel to the applied field ($||~\hat{\bm{x}}$).

\subsection{Power Dissipation: Dependence on Cylinder Radius}
\begin{figure}[tb]
\centering
\includegraphics[width=3.3in]{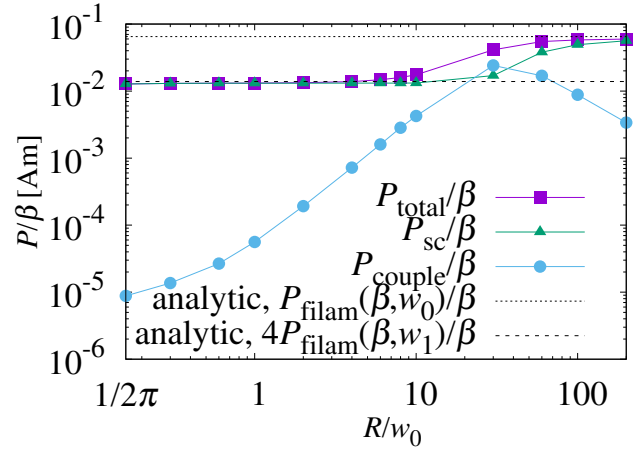}
\caption{Dependence of loss power per unit tape length on hollow-cylinder radius for $L_{\rm p}=6$~mm.
$P_{\rm total}$, $P_{\rm sc}$, and $P_{\rm couple}$ represent the total, SC, and coupling loss power, respectively.
The field sweep rate is fixed to $\beta=300$~mT/s.
The dotted and dashed lines show the loss power evaluated analytically by $P_{\rm filam}(\beta,w_0)/\beta$ and $4P_{\rm filam}(\beta,w_1)/\beta$ [Eq.~(\ref{loss-analytic})], respectively.
}
\label{fig2}
\end{figure}
The power dissipated on the tape surface per unit length is given by \cite{higashi2019_sust,higashi2020}
\begin{eqnarray}
P_{\rm total}&=&P_{\rm sc}+P_{\rm couple},\\
P_{\rm sc}&=&\int\int_{\rm filaments} {\rm d}\xi{\rm\rm d}\zeta p(\xi,\zeta),\\
P_{\rm couple}&=&\int\int_{\rm slots} {\rm d}\xi{\rm d}\zeta p(\xi,\zeta),\\
p(\xi,\zeta)&=&\frac{d_0}{L_{\rm tape}} \rho(\xi,\zeta)\left[ \left( \frac{\partial g}{\partial \xi}\right)^2+\left(\frac{\partial g}{\partial \zeta} \right)^2 \right].
\end{eqnarray}

Fig.~\ref{fig2} shows the cylinder-radius dependence of the loss power relative to $\beta$ for $L_{\rm p}=6$~mm and $\beta=300$~mT/s.
The total loss begins to increase at $R/w_0\sim 10$.
Even when the pitch length is a few times the tape width,
for $R\gtrsim 10 w_0$ in a rapidly ramped magnetic field with $\beta=300$~mT/s,
the SC filaments are coupled electrically with each other and the striation is ineffective.
However, in a rapidly ramped field, we can suppress the EM coupling by winding the SC tape with a thin round core up to $R/w_0\sim 1$. 
Therefore, in the CORC {\it wire} region (i.e., $R/w_0\sim 1$),
striation is effective for reducing the total loss.

For an effective tape width $w$,
by taking the thin-filament limit of $w/R\rightarrow 0$,
we obtain the following analytical formula for the loss power per unit length \cite{higashi2020},
which coincides with that for a twisted tape \cite{higashi2019}:
\begin{equation}
P_{\rm filam}\left(\beta,w\right)=\frac{B(\frac{2n+1}{2n},\frac{1}{2})}{\pi} \left( \frac{\beta w}{2 E_{\rm c}} \right)^{\frac{1}{n}}\frac{J_{\rm c}d_0 w^2 \beta}{2(2+1/n)},
\label{loss-analytic}
\end{equation}
where $B(p,q)=2\int_0^{\pi/2}{\rm d}\theta \cos^{2p-1}\theta\sin^{2q-1}\theta$ is the beta function with positive real numbers $p$ and $q$.

For $R/w_0\gtrsim 100$, the long $L_{\rm tape}$ means that 
the SC resistance $R^{||}_{\rm sc}$ becomes large relative to the transverse resistance $R_{\rm n}$ because $R^{||}_{\rm sc}\propto L_{\rm tape}$,
but $R_{\rm n}\propto L^{-1}_{\rm tape}$.
Therefore, the coupling current flows from the SC filaments to the resistive slots,
resulting in the EM coupling of the SC filaments.
In this case, we have $w\approx w_0$, and the analytical values of $P_{\rm filam}(\beta,w_0)/\beta$ agree well with the numerical ones (dotted line in Fig.~\ref{fig2}).
Here, we note that increasing the ratio of the round core radius to the total tape width, $R/w_0$, does not necessarily indicate the thick cable with $R\sim$ a few hundred millimeters.
Reducing $w_0$ also gives rise to an increase of $R/w_0$, although we fix $w_0$ and change $R$ in the present study.

Meanwhile, for $R/w_0\lesssim 1$,
$R^{||}_{\rm sc}$ remains small relative to $R_{\rm n}$.
Thus, no EM coupling occurs, and we have $w\approx w_1$.
The analytical values given by $4P_{\rm filam}(\beta,w_1)/\beta$ agree well with the numerical ones (dashed line in Fig.~\ref{fig2}); here, the factor 4 comes from the number of SC filaments.

\subsection{Power Dissipation: Dependence on Sweep Rate}
\begin{figure}[tb]
\centering
\includegraphics[width=3.5in]{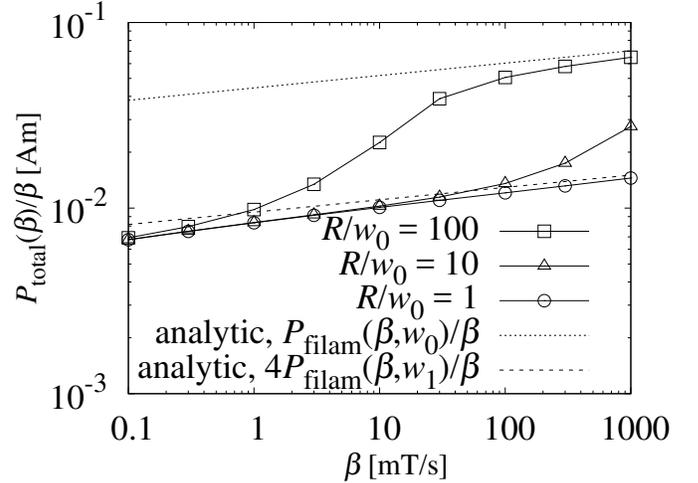}
\caption{Dependence of total loss power relative to $\beta$ on field sweep rate $\beta$ for $L_{\rm p}=6$~mm.
The dotted and dashed lines show the loss power evaluated analytically by $P_{\rm filam}(\beta,w_0)/\beta$ and $4P_{\rm filam}(\beta,w_1)/\beta$ [Eq.~(\ref{loss-analytic})], respectively.
}
\label{fig3}
\end{figure}
Fig.~\ref{fig3} shows how the total loss power relative to $\beta$ depends on the field sweep rate $\beta$ 
for $L_{\rm p}=6$~mm when $R$ is changed.
For $\beta_{\rm MRI}\sim 1$~mT/s, a typical sweep rate for an MRI magnet \cite{yokoyama2017,yachida2017},
there is no EM coupling for $R/w_0 \lesssim 10$.
However, in a rapidly swept field with $\beta \sim 100$~mT/s,
the striation is effective for reducing the loss only when $R/w_0\sim 1$.
For $R/w_0\gtrsim 10$,
the total loss increases slightly, meaning that the striation is ineffective even in the case of a short pitch of $L_{\rm p}/w_0=3$.

For large $\beta$,
$w_0$ is viewed as being the effective tape width because of the EM coupling.
Thus, in the case of $R/w_0=100$,
the loss power at $\beta=1000$~mT/s shows quantitatively good agreement with $P_{\rm filam}(\beta,w_0)/\beta$ (dotted line in Fig.~\ref{fig3}).
Meanwhile, for small $\beta$,
the effective tape width should be $w_1$,
and therefore $4P_{\rm filam}(\beta,w_1)/\beta$ agrees well with the numerics (dashed line in Fig.~\ref{fig3}) \cite{comment_small-value}.

\section{Discussion}
\begin{figure}[tb]
\centering
\includegraphics[width=3.5in]{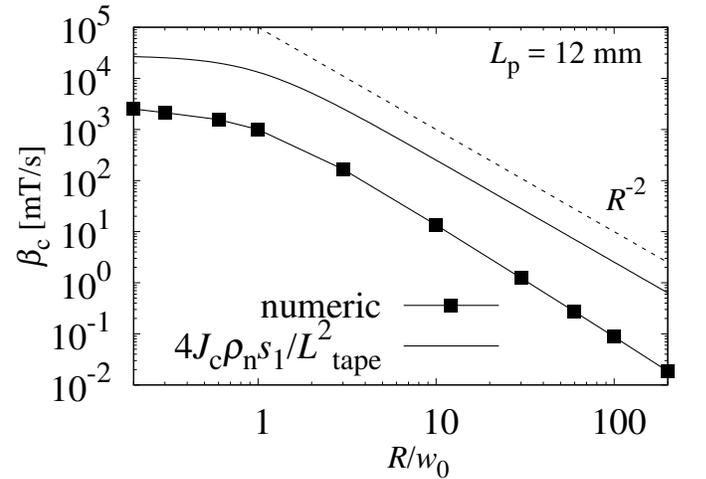}
\caption{Dependence of coupling sweep rate $\beta_{\rm c}$ on round-core radius $R$ for $L_{\rm p}=12$~mm.
The squares denote the numerical results, while the solid line with no symbols denotes the analytical result of Eq.~(\ref{beta_c-flat}).
}
\label{cylinder-radius-dep}
\end{figure}
\subsection{Electromagnetic coupling}
We examine the mechanism for EM coupling in a multi-filament helical tape
in the similar way as in Ref.\ \cite{higashi2019_sust}
by evaluating the coupling sweep rate $\beta_{\rm c}$, where the EM coupling starts.
Fig.~\ref{cylinder-radius-dep} shows how $\beta_{\rm c}$ depends on the hollow-cylinder radius for $L_{\rm p}=12$~mm.
Here, $\beta_{\rm c}$ is defined as $\beta$ satisfying 
$P_{\rm couple}(\beta)/P_{\rm sc}(\beta)\approx 0.02$,
and it is obtained numerically by subjecting $\beta$ to linear interpolation (the squares in Fig.~\ref{cylinder-radius-dep}).
The choice of $P_{\rm couple}/P_{\rm sc}$ affects the numerical results for $\beta_{\rm c}$
but is irrelevant for the $R$ dependence, which is our interest here.
We can understand how $\beta_{\rm c}$ depends on $R$ 
by evaluating $\beta_{\rm c}$ analytically from the EM coupling condition, namely, 
\begin{equation}
\frac{R^{||}_{\rm sc}}{R_{\rm n}}=\frac{\beta}{\beta_{\rm c}}> 1.
\end{equation}
Herein, the SC filament resistance in the longitudinal ($\zeta$) direction can be evaluated as
$R^{||}_{\rm sc}=\rho_{\rm sc}(L_{\rm tape}/2)/w_1 d_0$.

For a nearly flat tape with a thick round core with $R/w_0 \gg 1$,
the length scale at which magnetic flux vortices enter a superconductor can be viewed as being the filament width $w_1$,
and therefore the electric field in the superconductor is evaluated as $\sim\beta w_1$ (e.g., Ref.\ \cite{mawatari1997} for the disk geometry).
Consequently, the SC resistivity is evaluated as $\rho_{\rm sc}\sim \beta w_1/J_{\rm c}$.
The transverse normal resistance is $R_{\rm n}=\rho_{\rm n}s_1/(L_{\rm tape}/2)d_0$.
Therefore, we obtain the coupling sweep rate in the case of $R/w_0\gg1$ as
\begin{equation}
\beta_{\rm c} \sim \frac{4J_{\rm c}\rho_{\rm n}s_1}{L^2_{\rm tape}}\rightarrow \frac{J_{\rm c}\rho_{\rm n}s_1}{\pi^2 R^2}\propto R^{-2}~~(kR\gg 1).
\label{beta_c-flat}
\end{equation}

Meanwhile,
in the case of a nearly tubular wire with a thin round core ($R/w_0\sim 1$),
the characteristic length scale of the flux motion can be viewed as being the core radius $R$,
and therefore the electric field is evaluated as $\sim \beta R$,
leading to $\rho_{\rm sc}\sim \beta R/J_{\rm c}$.
Therefore, the coupling sweep rate in the case of $R/w_0\sim 1$ is estimated as
\begin{equation}
\beta_{\rm c} \sim \frac{4J_{\rm c}\rho_{\rm n}s_1}{L^2_{\rm tape}}\frac{w_1}{R} \rightarrow \frac{J_{\rm c} \rho_{\rm n}s_1 w_1}{\pi^2R^3}\propto R^{-3}~~(kR\gg 1).
\label{beta_c-tubular}
\end{equation}

In Fig.~\ref{cylinder-radius-dep}, the solid line with no symbols corresponds to $\beta_{\rm c}$ evaluated analytically by Eq.~(\ref{beta_c-flat}).
Here, $\beta_{\rm c}$ scales as $R^{-2}$ for $R/w_0\gg 1$, and its dependence on $R$ agrees well with the numerics in the whole range of $R$.
In the case of the nearly tubular wire with $R/w_0\sim 1$,
as estimated by Eq.~(\ref{beta_c-tubular}) for $R/w_0\gg1$,
$\beta_{\rm c}$ scales as $R^{-3}$,
in disagreement with the numerics.
Based on how $\beta_{\rm c}$ depends on $R$,
we conclude that the EM coupling of a multi-filament helical tape reflects the penetration of magnetic flux from the long edges of the SC filaments.

\subsection{Relation to coupling time constant}
The measurement of ac losses for a fixed low amplitude of a sinusoidally oscillating applied magnetic field
with sweeping a frequency $f$ as in Refs. \cite{li2020,sogabe2020} could be comparable to our calculations
in which an applied field is actually swept at a constant rate $\beta$.
Note that ac loss measurements for a fixed $f$ with sweeping an applied field as in Ref. [6] is never comparable to our simulations.
Alternating-current losses at low fields and high frequencies are dominated by coupling losses because the energy dissipation inside superconductors is negligible at sufficiently low fields.
In the case of a sinusoidally oscillating magnetic field, the coupling time constant is given by $\tau_{\rm c}=1/2\pi f_{\rm c}$ with the coupling frequency $f_{\rm c}$.
In this case, the coupling field sweep rate can be viewed as $2\pi f_{\rm c}B$, which is corresponding to $\beta_{\rm c}$ in our notation; $\beta_{\rm c}\equiv 2\pi f_{\rm c} B$.
Here, note that $B$ denotes the magnetic field on the SC tape.
Therefore, in the case of $R/w_0\gg 1$ by using the estimation of $\beta_{\rm c}$ (\ref{beta_c-flat}),
we may relate the coupling time constant to the round core radius as
\begin{equation}
\tau_{\rm c}\sim\frac{\mu_0 \left[ L^2_{\rm p} +(2\pi R)^2\right]}{4\pi \rho_{\rm n}}\left( \frac{d_0}{s_1}\right),
\label{coupling-time-constant}
\end{equation}
where $\mu_0$ the vacuum permeability and
the magnetic field on the SC tape is evaluated as $B\sim\mu_0 J_{\rm c} d_0/\pi$ based on the Bean model.
The estimation (\ref{coupling-time-constant}) can be obtained via the same way as in Ref. \cite{higashi2019_sust} in the case of an alternating-current field.  
One can immediately see from the relation (\ref{coupling-time-constant}) that 
with increasing $R$ the time scale for a decaying coupling current becomes long, meaning that EM coupling starts from at low frequencies.

\subsection{Effects of multi layers and tapes}
We consider that SC tapes are subjected to an applied magnetic field as high as a few tesla, which is far above a full flux penetration field $B_{\rm p}$.
For concreteness, $B_{\rm p}$ for striated helical tapes is approximately estimated via the flat tape formula for the filament width $w_1$;
$B_{\rm p} = (\mu_0J_{\rm c}d_0/\pi)[1 + \ln(w_1/d_0)] \approx 0.260$ T \cite{brandt1996}.
In the case of the multi-layer configuration, $B_{\rm p}$ will be a few times more than that for a single flat tape, but it is still much smaller than a few tesla. 
Accordingly, the magnetic flux penetrates the tapes completely.
In this condition, because the response to an applied field is dominant, the interaction between SC tapes is relatively small and negligible.
Therefore, the EM coupling among SC filaments in the tapes is not seriously influenced by the interaction between the striated SC tapes in the different layers.  
However, at low magnetic fields, the interaction between the tapes in the different layers might be important to model a CORC wire wound by several tapes and layers \cite{majoros2014,wang2019}.

\section{Summary}
Regarding the excitation and demagnetization of an MRI magnet, 
we simulated numerically the magnetization loss of a multi-filament helical SC tape in the steady state in a constantly ramped magnetic field.
Even in a rapidly ramped field with a rate of $\beta=300$~mT/s,
no EM coupling of the SC filaments was seen for $R/w_0\sim 1$ because of the short pitch length.
However, for $R\gtrsim 10w_0$, the SC filaments were coupled electrically,
showing the striation to be ineffective.
The coupling sweep rate scaled as $R^{-2}$ for $R/w_0\gtrsim 1$,
reflecting the penetration of magnetic flux from the tape edges.
This behavior is considered to be similar to the EM response of a flat tape rather than that of a tubular wire, 
 even when the round core is as narrow as the tape width, which is nearly a tubular wire.

\section*{Acknowledgments}

We thank M.\ Vojen{\v{c}}iak (Slovak Academy of Sciences) for discussions during the 31st International Symposium on Superconductivity (ISS2018).
We also thank N.\ Amemiya (Kyoto University) and Y.\ Yoshida (AIST) for providing us with the pre-publication manuscript of Ref.\ \cite{li2020}.
The present work is based on results obtained from a project commissioned by the New Energy and Industrial Technology Development Organization (NEDO).

%


\ifCLASSOPTIONcaptionsoff
  \newpage
\fi

\bibliographystyle{IEEEtran}
\bibliography{IEEEabrv,IEEEexample}

\begin{thebibliography}{10}
\bibitem{comment_small-value}
The simulated losses are smaller than the analytically evaluated values. This
  is because we put the numerical value in the current density to avoid
  numerical instabilities in the following way;
  $|\bm{J}(\xi,\zeta)|^2=(J^2_\xi+J^2_\zeta)/J^2_{\rm c}+\delta$ with
  $\delta=10^{-1}$. However, the choice of $\delta=10^{-1}$ results in somewhat
  smaller numerical value than analytically expected one; that is,
  magnetization losses are numerically underestimated. Especially, effect of
  $\delta$ becomes prominent when the current is small at low sweep rates
  $\beta$, and thereby this underestimation is pronounced. On the other hand,
  at high $\beta$, because the current is large, the $\delta$ in the current
  has relatively low impact. Thus the discrepancy between the numerical value
  and the analytical one becomes small as seen in Fig. \ref{fig3} (Compare the
  discrepancy at high $\beta$ with that at low $\beta$). Indeed, we confirmed
  the discrepancy tends to be somewhat small for $\delta=0.6\times10^{-1}$, but
  the dependence of the loss power on $\beta$ does not change qualitatively and
  our conclusion is not altered.
\end{thebibliography}
\end{document}